\documentclass[aps,11pt]{revtex4-1}
\usepackage[top=1in, bottom=1.5in, left=1.5in, right=1.5in]{geometry}  

\usepackage{amsmath}
\usepackage{amssymb}
\usepackage{graphicx}
\usepackage{dcolumn}
\usepackage{bm}
\usepackage[english]{babel}
\usepackage{setspace}			
\usepackage{hyperref}         

\hypersetup{colorlinks=true}  

\makeatletter
\let\reftagform@=\tagform@
\def\tagform@#1{\maketag@@@{\color{blue}(\ignorespaces{#1}\unskip\@@italiccorr)}}
\renewcommand{\eqref}[1]{\textup{\reftagform@{\ref{#1}}}}
\makeatother

\usepackage{color} 
\definecolor{PaleYellow1}{rgb}{1.0,1.0,0.5}
\definecolor{Yellow1}{rgb}{1.0,1.0,0.5}
\definecolor{LinkColor}{rgb}{0.0,0.45,0.0} 
\definecolor{Grey0}{rgb}{0.95,0.95,0.95} 
\definecolor{Grey1}{rgb}{0.92,0.92,0.92} 
\definecolor{Grey2}{rgb}{0.4,0.4,0.4} 
\definecolor{Green1}{rgb}{0.4,0.9,0.4}
\definecolor{Orange1}{rgb}{1.0,0.7,0.05}
\definecolor{DarkBlue}{rgb}{0,0.08,0.45}
\definecolor{DarkRed}{rgb}{0.75,0.08,0.0}
\definecolor{BrightGrey}{rgb}{0.4,0.4,0.4}
\definecolor{BrightGreen}{rgb}{0.1,0.8,0.1}
\definecolor{Orange}{rgb}{1.0,0.5,0.01}
\definecolor{BlueGreen}{RGB}{12,201,179}
\definecolor{kbficolor}{RGB}{171,83,83}
\definecolor{etiscolor}{RGB}{81,106,156}

\hypersetup{
    colorlinks=true,
    linkcolor=blue,                  
    filecolor=red,   
    urlcolor=blue,                   
    citecolor=LinkColor,             
    anchorcolor=Grey2,
    pdftitle={M.Patriarca, Antti Kuronen, and Kimmo Kaski - Nucleation and dynamics of dislocations in mismatched heterostructures}, 
    bookmarks=true,
    pdfpagemode=FullScreen,
}

    \makeatletter
    \renewcommand*{\@fnsymbol}[1]{\ensuremath{\ifcase#1
      \or {\color{DarkRed}  (*)}
      \or {\color{DarkRed} (1)}
      \or {\color{DarkRed} (2)}
      \or {\color{DarkRed} (3)}
      \or {\color{DarkRed}  (**)}
      \or \ddagger
      \or \mathsection
      \or \mathparagraph
      \or \dagger\dagger
      \or \ddagger\ddagger
      \else\@ctrerr\fi}}
    \makeatother
 
\begin{document}

\setstretch{1.1}

\title{Nucleation and dynamics of dislocations\\in mismatched heterostructures }
 \thanks{\small
	\noindent
	Presented at the {\it 2001 \href{https://www.mrs.org/}{MRS} Fall Meeting}, Boston, MA, November 26-29, 2001.
	\vspace{0.1cm} \\ \noindent
    This is the draft of the paper eventually published as:
    Marco Patriarca, Antti Kuronen, and Kimmo Kaski,  {\it Nucleation and dynamics of dislocations in mismatched heterostructures},  
	in: ``Current Issues in Heteroepitaxial Growth-Stress Relaxation and Self Assembly'',
	2001 Materials Research Society Symposium Proceedings, Symposium N 4.4, Vol. 69, 2002.   
	E.A. Stach \emph{et al.}. Editors. ~ {\bf doi}:~\href{https://doi.org/10.1557/PROC-696-N4.4}{{10.1557/PROC-696-N4.4}}
	\vspace{0.1cm} \\ \noindent
	The present version of the article contains links to online videos of some numerical simulations discussed in the article. Such videos are available, together with other related videos, at \href{https://sites.google.com/view/mcp-defects}{{\tt  https://sites.google.com/view/mcp-defects}}
	\vspace{0.1cm} \\ \noindent
	M. Patriarca acknowledges support from the European Regional Development Fund (ERDF) Center of Excellence (CoE) program grant TK133, which made the realization of this electronic version possible.\\
}

\author{
Marco Patriarca
}
\thanks{\small
		Current Affiliation:
		{\it NICPB--National Institute of Chemical Physics and Biophysics, R\"avala 10, 10143 Tallinn, Estonia}.
		~~
		E-mail:~{{\tt marco.patriarca}\,@\,{\tt kbfi.ee }}
}

\author{
Antti Kuronen
}
\thanks{\small
		Current Affiliation:
		{\it Department of Physics, Accelerator Laboratory, P.O.Box 43, FIN-00014 University of Helsinki, Finland}.
		~~
		E-mail:~{{\tt antti.kuronen}\,@\,{\tt helsinki.fi }}
}

\author{
Kimmo Kaski
}
\thanks{\small
		Current Affiliation:
		{\it Department of Computer Science, Aalto University School of Science, P.O.Box 15500, Fl-00076 Aalto, Finland}.
		~~
		E-mail:~{{\tt kimmo.kaski}\,@\,{\tt aalto.fi }}
}

\affiliation{	
	Helsinki University of Technology, Laboratory of Computational Engineering, P.O.Box 9400, FIN-02015 HUT, FINLAND
}\thanks{\small
	Original affiliation. Current affiliations and contact information in notes {\color{DarkRed}  (1)}-{\color{DarkRed}  (3)} above. 
}




\begin{abstract}
\vspace{1.0cm}
\normalsize
\noindent
{\bf Abstract}.~
In this paper we have investigated, through computer simulations, dislocation nucleation
and dislocation dynamics in a heterostructure system with the lattice-mismatch interface,
i.e. a system with internal strain.
In particular, we have studied the dependence of the
nucleation thresholds on the basic parameters of the crystals, such as the amount of
mismatch and the system temperature.
These studies have been carried out by using the
simulation code with a graphical user interface developed at our laboratory.
This on-line
simulation system produces a real time interactive visualization of the 3-D Molecular
Dynamics model.
Furthermore, it detects the presence of dislocations and tracks them
by an algorithm based on potential energy mapping.
\end{abstract}

\maketitle

\section{Introduction}

Recent studies of nucleation of dislocations and dislocation dynamics in lattice-mismatched
heterostructures have resulted in a lot of interest due to the technological importance and
applications of such structures [1, 2]. In our previous studies we have investigated the
interaction between dislocations and a misfit interface in a two-dimensional Lennard-Jones
system [3, 4]. In these studies we observed reactions between dislocations occurring
by creation of partial dislocations and a stacking fault. These reactions enabled the
dislocations to migrate to the interface and relieve misfit stress.

Our previous studies of
dislocation nucleation in a two-dimensional lattice-mismatched system [5] indicate that
nucleation is clearly a thermally activated process and that it is asymmetric with respect
to the sign of the mismatch. The present work is an extension of these two-dimensional
studies to three-dimensional lattice-mismatched structures. We present some numerical
results concerning the study of dislocation nucleation at the mismatch interface between
two different materials, only subject to the internal strain arising from the mismatch.

\begin{figure}[hb!]
	\centering
	\includegraphics[width=10cm]{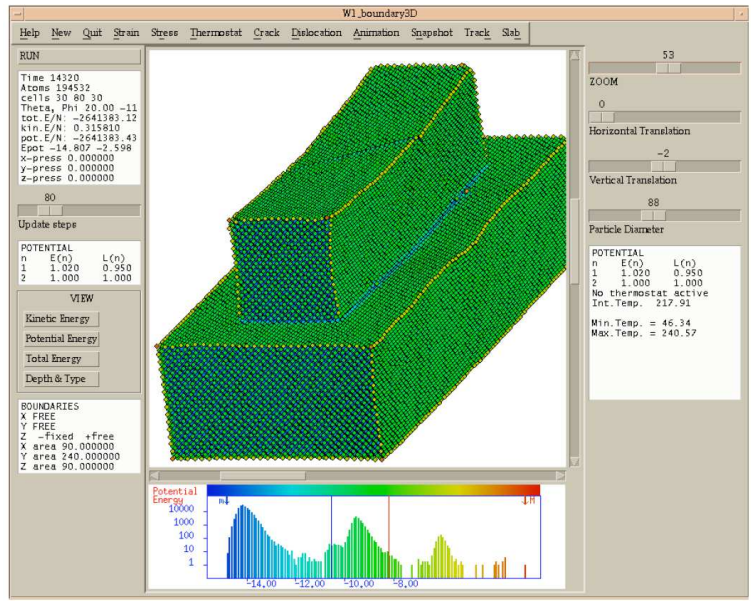}
	\caption{
		Visualization of the dot-sample considered in this paper with the Graphical User Interface of the code {\it Boundary3D}, in which different types and scales of visualization of the atoms can be chosen in real time while the simulation run is executed. Here colors in the main window represent mean potential energies of single atoms, while the lower plot shows the corresponding histogram in potential energy and the energy-color correspondence.
		Different colors can be approximately put into correspondence with the atoms in different parts of the sample: bulk - blue, surface - green, edges - yellow, corners - red.
		\label{fig-1a}
	}
\end{figure}

\section{Model}

Our idealized model of a system with misfit interface consists of two fcc crystals with
different lattice constants, joined together. The interface is in the (001) plane. Here
we concentrate on the representative example of a quantum dot geometry. The initial
conditions were prepared as an assembly of two different types of atoms located on a
regular {\it fcc} lattice with a unique lattice constant. The sample size is $80 \times 30 \times 30$ lattice
constants.
The atoms of type 1 (2 ) were located in the upper dot (lower substrate) of the sample.
We limited ourselves to systems without periodic boundary conditions.
This choice was made in order to avoid the presence of unphysical interactions and constraints,
which could strongly influence the behavior of the system, because of the introduction of the following two artificial length scales:\\
-- The first length scale is the fixed lattice constant. The actual lattice constant in general depends on
the thermal status of the system and should not be imposed by assigning a certain system size and periodic boundary conditions.
If periodic boundary conditions are necessary, the lattice constant could be suitably determined e.g. by a constant
pressure molecular dynamics algorithm.
A dynamics modified  by such constraints  on a time scale comparable or larger than that of dislocation dynamics may strongly affect
the phenomena under study.\\
-- The other artificial length scale is the linear size $L$ of the system along the periodic dimension, which represents a constraint on the density of dislocations, unless the number of dislocations present in a region of size $L$ is much larger than one,
which in turn requires the model system to have a very large number of atoms.

As for the system size, there is both an upper and a lower limit.
On the one hand, this study is based on classical molecular dynamics simulations, for which the system size
is a practical issue,  because of the computing time. Here we have limited ourselves to
sample sizes up to $2 \times 10^5$ atoms.
On the other hand, there is a lower size limit at which dislocation nucleation can take place: if the system is too small, the strain is not sufficient to induce the nucleation of a dislocation.
For instance, one of the linear system sizes must be at least of the order of 100 lattice constants, for a 5\% lattice misfit, in order for at least one dislocation to appear.

Since our MD simulations are exploratory in nature, we will use simple pairwise Lennard-Jones
potentials for simulating the materials, i.e., a potential $V_{LJ}(x; \sigma_1, \epsilon_1)$ and
a potential $V_{LJ}(x; \sigma_2, \epsilon_2)$ for the interaction between atoms of the same type, 1 or 2, respectively.
For simplicity, we set the well depth equal in both materials, $\epsilon_1 = \epsilon_2$, while the length parameters are different,
$\sigma_1 \ne \sigma_2$.
The misfit is expressed in terms of the two different lattice constants $a_1$ and $a_2$ as $m = (a_1 - a_2)/a_2$.
The potential was cut-off at about two lattice constants, so that every atom of the system interacts with
the first two shells of nearest neighbors. The interaction between two atoms of different
types 1 and 2 (i.e. across the misfit interface) is described with a potential of the same Lennard-Jones form,
with parameters $\epsilon$ and $\sigma$ determined through a standard interpolation procedure, see Ref. [3] for details.

\begin{figure}[ht!]
	\centering
	\includegraphics[width=10cm]{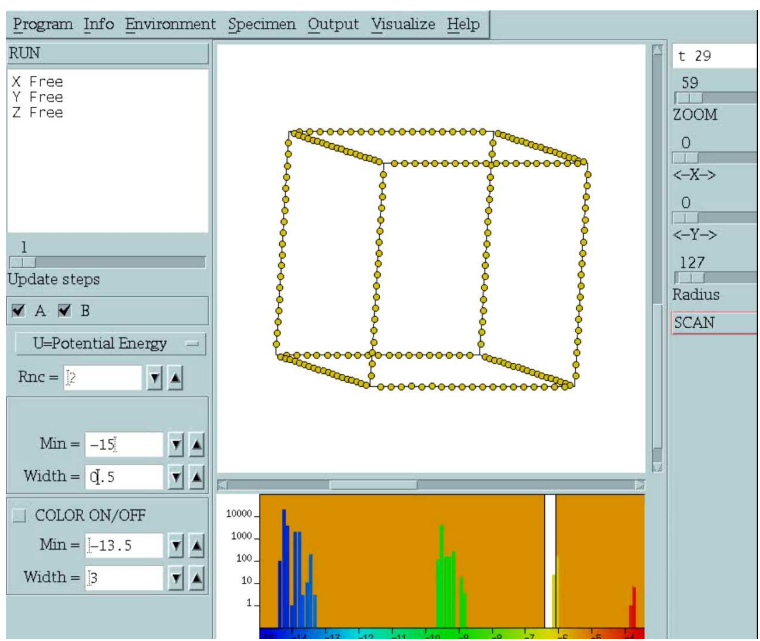}
	\caption{
		Selected visualization of a simple cubic structure of Lennard-Jones atoms, obtained by filtering atoms according to their mean potential energy.
		The potential energy window selected here corresponds to the edges of the cubic structure.
		A video showing the selected visualization corresponding to a moving potential energy window scanning the energy range from the lowest energies (bulk) to the highest energies (corners) can be seen at  \href{https://youtu.be/GMjkL8Yi-rE}{{\tt  https://youtu.be/GMjkL8Yi-rE}}
		\label{fig-1b}
	}
\end{figure}

In dislocation nucleation and dynamics, the temperature $T$ of the system plays an important role.
We evaluate it as usually from the average kinetic energy.
In order to have a feeling of the realistic values of the temperature range involved,
we have assigned values to the potential parameters of the interatomic potential of the substrate
material corresponding to copper.

Another important issue of our MD simulations is the choice of thermostat. We have found
that the use of a Langevin or a Nose-Hoover type thermostat, for keeping the system at
a constant temperature, may strongly influence the dynamics of the system, even to the
point of eventually preventing any dislocation nucleations. For this reason the use of a
(Langevin) thermostat has been limited to the very initial phase of the simulation. This
was done in order to prevent the excessive heating of the sample, which would be caused
by an initial configuration with all atoms forced on the same lattice. The temperature of the
sample may undergo some variations during the simulation run, but then relaxes, on a time
scale of the order of 1 nanosecond, toward a temperature which is assumed to represent the
final equilibrium temperature of the system. All the numerical simulation runs were carried
out, for most of the time steps, by using standard unconstrained molecular dynamics.

\section{Simulation Tools}

The numerical studies have been carried out by using our newly developed graphical
user interface simulation code ``Boundary3D''. This code is based on a 2D version, also
developed by some us [6], already used to study the dynamics of dislocations in 2D samples
[3–5]. This on-line simulation system allows a real time interactive visualization of a
Molecular Dynamics model for crystalline systems. It also allows the user to vary the type
of visualization as well as the main parameters of the system, see Fig.~\ref{fig-1a}.
The graphical user interface is particularly important because of its ability to visualize dislocation dynamics in
real time and to track their movement through the system. It turned out that the tracking is
realizable through a mapping, which is based on a suitable microscopic quantity, such as the
effective single particle potential energy.

\begin{figure}[ht!]
	\centering
	\includegraphics[width=8cm]{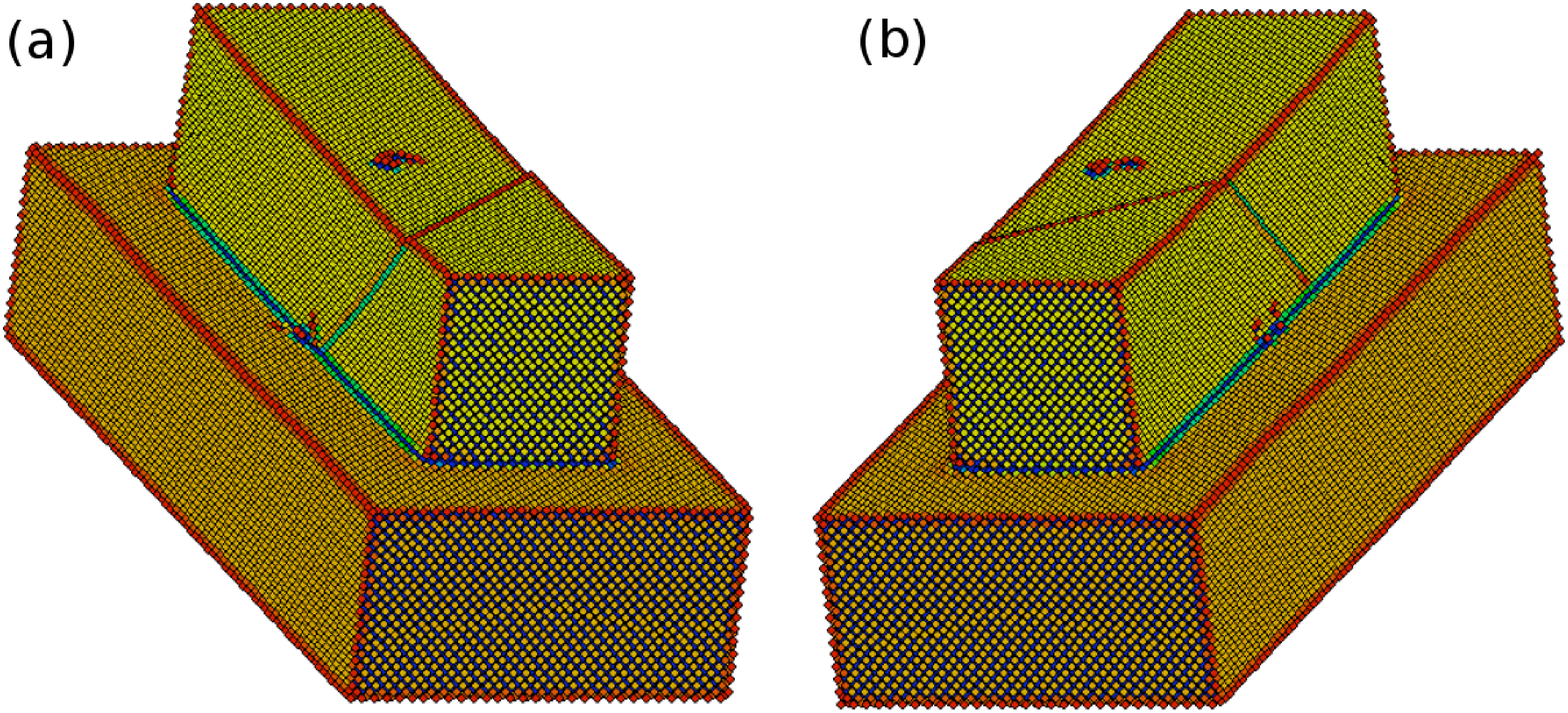}
	\includegraphics[width=3.5cm]{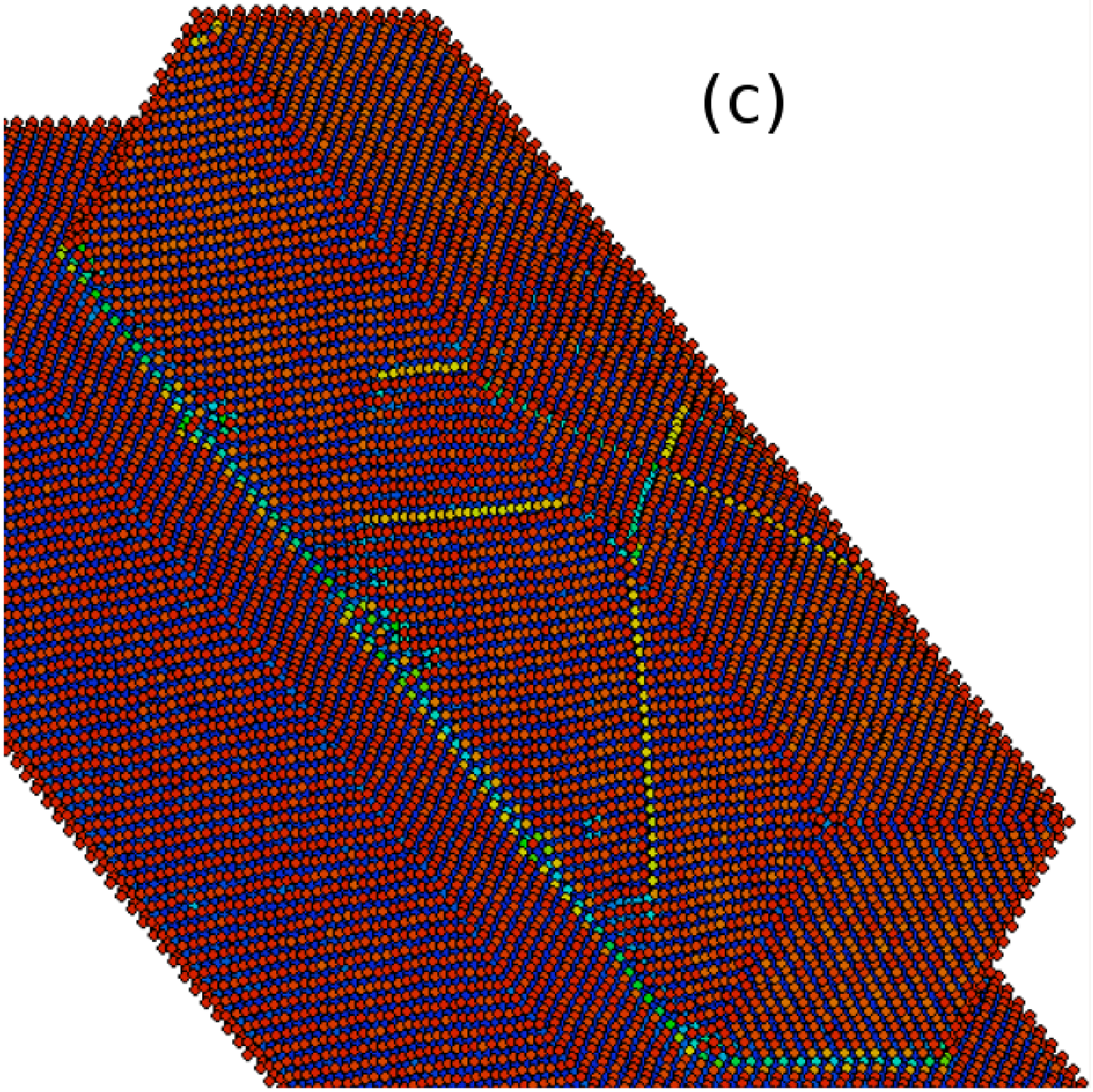}
	\caption{
		(a)-(b): Left and right view of a sample with dot geometry at $T = 200$ K and misfit $m = 5\%$.
		(c) Same Sample at temperature $T = 240$.
		The color of the atoms is coded according to potential energy (different color scales are used for a better visualization).
		The lines visible on the sides of the dot are the steps associated to the stacking faults and are characterized by a higher potential energy.
		\label{fig-2}
	}
\end{figure}

In order to illustrate the physical principle of how
the tracking procedure works, it is worth considering Fig.~\ref{fig-1a}, in which the aspect of the main
interface of the code is shown. In the upper graphical box a sample with a quantum dot
geometry is drawn, and atoms are colored according to their effective potential energies.
The mapping between color and potential energy is represented in the lower graphical
box, together with the potential energy distribution.
It can be seen that the potential energy histogram consists of four peaks, which can be clearly distinguished from each other and visualized with different colors.
From left to right these peaks correspond to responses from the bulk,
surface, border, and corner atoms, respectively. On the other hand, it should noted that in a
perfect (infinite) crystal there would appear only a single peak, corresponding to the bulk atoms.
Thus,
these boundaries, i.e. surfaces, borders, and corners, represent
particular types of crystal defects with different values of the atomic average potential energies
of the system.

Using the same criterion, one can also scan the system under study in potential energy to reveal selectively atoms corresponding to one of the peaks, as shown in Fig.~\ref{fig-1b} in the case of a cubic sample of Lennard-Jones atoms for the sake of clarity --- see also the associated video at \href{https://youtu.be/GMjkL8Yi-rE}{{\tt  https://youtu.be/GMjkL8Yi-rE}}

These simple considerations can be generalized in order to visualize quite
arbitrary types of crystal defects, such as dislocations or point defects.

\section{Results}

The results presented below are based on a set of simulations, carried out for about $2 \times 10^4$
time steps, corresponding to 20 ps for standard values of the other parameters. For a
given amount of mismatch and size of the sample, we have found that there are three
regions, in the temperature scale, in which the sample shows a different behavior. In the
low temperature region, i.e. $T < 200$ K and a 5 \% misfit, no dislocation nucleation was
observed. Eventually, only the deformation due to the misfit stress was visible. Sometimes,
it was also possible to observe the initial phase of a unstable dislocation nucleation,
followed by its reabsorption. On the other hand, in the intermediate temperature region, i.e.
$200~K < T < 300~K$, and a 5\% misfit, the nucleation of one or more dislocations is clearly
visible and the dislocations remain stable with time.

\begin{figure}[ht!]
	\centering
	\includegraphics[width=10cm]{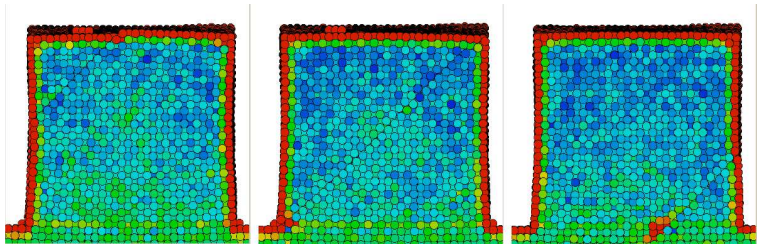}
	\caption{
		Slabs taken at different depths of the sample of Fig.~\ref{fig-2}-(a), showing how the  (111) partial dislocation cuts through the sample.
		\label{fig-3}
	}
\end{figure}

We have observed the formation of partial dislocations, which cut the dot along the inclined (111) planes.
Figs.~\ref{fig-2}-(a) and \ref{fig-2}-(b) show a single dislocation nucleated in a sample with 5 \% misfit and  $T = 200~K$.
With increasing temperature, the number of dislocations increases, but not more than 2-3
stable dislocations were observed, probably because of the small size of the sample.
In the snapshot in Fig.~\ref{fig-2}-(c), obtained at temperature $T = 240$ K, one can distinguish three partial dislocations instead of
only one.

\begin{figure}[ht!]
	\centering
	\includegraphics[width=10cm]{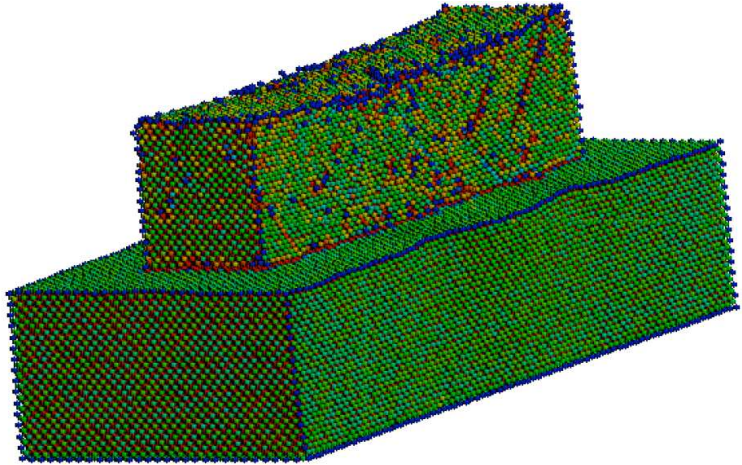}
	\caption{
		A sample with 5 \% misfit at $T = 300$ K.
		Defects on the dot sides in different colors are due to short-lived dislocations.
		See also the related video at \href{https://youtu.be/vmeBHDKzEeY}{{\tt  https://youtu.be/vmeBHDKzEeY}}
		\label{fig-4}
	}
\end{figure}

We have also studied the structure of the dislocation in the inner bulk, by taking a
cross section of the the sample.
In Fig.~\ref{fig-3} we present a successive sequence of cross sections
taken from the sample of Fig.~\ref{fig-2}-(a) in the (010) plane.
If the temperature is further increased, i.e.  at $T > 300$ K, with a 5 \% misfit, besides the usual - stable dislocations one can also
observe some other short-lived (unstable) dislocations, which through nucleation appear
and disappear in the crystal structure on a time scale of few time steps,
as in the snapshot in Fig.~\ref{fig-4} (see also the related video at \href{https://youtu.be/vmeBHDKzEeY}{{\tt  https://youtu.be/vmeBHDKzEeY}})

It should be mentioned that these temperature thresholds for dislocation nucleation
depend on the amount of misfit. In addition, our simulations indicate that the temperature
threshold decreases when the amount of misfit is increased.

\section{Conclusions}

In this study we have investigated dislocation nucleation in heterostructures with a
lattice-mismatch by using a simulation code with a graphical user interface. Despite the fact
that the nucleation of a dislocation is a statistical event, our preliminary simulation results
indicate a definite relation between misfit and the critical temperature at which dislocations
begin to appear. Further studies are under way in order to determine such a relation more
precisely.

\vspace{1cm}

\begin{acknowledgments}
Work supported by the Academy of Finland, Research Centre for Computational Science and Engineering, project no. 44897 (Finnish Centre of Excellence Program 2000-2005).
\end{acknowledgments}

\section*{References}

\noindent
[1] E. A. Fitzgerald, {\it Dislocations in strained-layer epitaxy: theory, experiment, and applications}, Mat. Sci. Rep. {\bf 7}(3), 87-142 (1992).\\

\noindent
[2] R. Hull and J. C. Bean, {\it Misfit dislocations in lattice-mismatched epitaxial films}, Crit. Rev. Solid State Mat. Sci. {\bf 17}(8), 507-546 (1992).\\

\noindent
[3] A. Kuronen, K. Kaski, L. Perondi, and J. Rintala, {\it Atomistic modelling of interaction between dislocations and misfit interface}, Europhys. Lett. {\bf 55}, 19 (2001).\\

\noindent
[4] A. Kuronen, K. Kaski, L. F. Perondi, and J. Rintala, {\it Interaction Between Dislocations and Misfit Interface},
in Materials Research Society Symposium Proceedings,  Vol. 634, B4.9.1,
Symposium B ``Structure and Mechanical Properties of Nanophase Materials -- Theory \& Computer Simulation \emph{vs} Experiment''.\\

\noindent
[5] K. Kaski, A. Kuronen, and M. Robles, {\it Dynamics of Dislocations in a Two-dimensional System}, p. 12, 
in ``Computer Simulation Studies in Condensed Matter Physics XIV'',
D. P. Landau, S. P. Lewis, and H. B. Sch\"uttler, editors (Heidelberg, Berlin, 2001, Springer Verlag).\\

\noindent
[6] J. Merimaa, L. Perondi, and K. Kaski, {\it An interactive simulation program for visualizing complex phenomena in solids}, Comp. Phys. Comm. {\bf 124}, 60-75 (1999).

\end{document}